\pdfoutput=1

\documentclass[preprints,article,accept,moreauthors,pdftex,10pt,a4paper]{mdpi}
\firstpage{1} 
\pubvolume{xx}
\issuenum{1}
\articlenumber{1}
\pubyear{2018}
\copyrightyear{2018}
\externaleditor{Academic Editor: name}
\history{Received: date; Accepted: date; Published: date}

\usepackage{pgfplots} 
\pgfplotsset{compat=1.14} 
\usepgflibrary{shapes.misc} 
\usepackage{tikz}
\usetikzlibrary{arrows, shapes,external, decorations.pathmorphing,
    decorations.pathreplacing, backgrounds, positioning, fit, pgfplots.ternary, pgfplots.units}
\usetikzlibrary{math}

\PassOptionsToPackage{inline}{enumitem}

\usepackage{amsfonts}
\usepackage{amssymb} 
\usepackage{mathtools}
\renewcommand{\intertext}[1]{\shortintertext{#1}}

  \newtheorem{cons}{Construction}

\RequirePackage{etoolbox}
\providetoggle{extendedVersion}
\toggletrue{extendedVersion}
\togglefalse{extendedVersion}

\providetoggle{ieeeVersion}
\toggletrue{ieeeVersion}
\providetoggle{graphicalSummary}
\toggletrue{graphicalSummary}
\togglefalse{graphicalSummary}

\graphicspath{{../figs/}{../pics/}}
\usepackage{subcaption}

\usepackage{url}

\usepackage{pgfplots} 
\pgfplotsset{compat=1.14} 
\usepgflibrary{shapes.misc} 
\RequirePackage{tikz}
\usetikzlibrary{arrows, shapes,external, decorations.pathmorphing,
    decorations.pathreplacing, backgrounds, positioning, fit, pgfplots.ternary, pgfplots.units}

\usepackage{tabularx} 
\usepackage{multirow} 

\usepackage{xargs}                      
\usepackage[colorinlistoftodos,prependcaption,textsize=tiny]{todonotes}
\newcommandx{\unsure}[2][1=]{\todo[linecolor=red,backgroundcolor=red!25,bordercolor=red,#1]{#2}}
\newcommandx{\change}[2][1=]{\todo[linecolor=blue,backgroundcolor=blue!25,bordercolor=blue,#1]{#2}}
\newcommandx{\info}[2][1=]{\todo[linecolor=OliveGreen,backgroundcolor=OliveGreen!25,bordercolor=OliveGreen,#1]{#2}}
\newcommandx{\improvement}[2][1=]{\todo[linecolor=Plum,backgroundcolor=Plum!25,bordercolor=Plum,#1]{#2}}
\newcommandx{\thiswillnotshow}[2][1=]{\todo[disable,#1]{#2}}

\newcommand{\supp}[1]{\ensuremath\operatorname{supp}\left (#1\right)} 
\newcommand{\dual}[1]{\ensuremath#1^\delta} 

\newcommand{\xbot}[1][\vec x]{\ensuremath{B_{#1}}}
\newcommand{\xtop}[1][\vec x]{\ensuremath{T_{#1}}}
\newcommand{\xfin}[1][\vec x]{\ensuremath{F_{#1}}}

\input{maxplus_minplus.tex}
\input{Galois_connections.tex}
\newcommand{\hmean}[3][r]{\ensuremath{M_{#1}(#2,#3)}}
\newcommand{\shmean}[3][r]{\ensuremath{\hat M_{#1}(#2,#3)}}
\newcommand{\pentropy}[2][r]{\ensuremath{\tilde H_{#1}(#2)}}

\newcommand{\hinf}[1]{\ensuremath{\mathfrak I_\ast\left(#1\right) }}
\newcommand{\hinfinv}[1]{\ensuremath{\left (\mathfrak I_\ast\right)^{-1}\left(#1\right)}}

\newcommand{\rentropy}[2][\alpha]{\ensuremath{H_{#1}(#2)}}

\newcommand{\rdist}[2][\alpha]{\ensuremath{q_{#1}(#2)}} 

\newcommand{\srprob}[2][r]{\ensuremath{\tilde P_{#1} \left (#2\right)}}
\newcommand{\srpot}[2][r]{\ensuremath{\tilde V_{#1} \left (#2\right)}}
\newcommand{\srentropy}[2][r]{\ensuremath{\tilde H_{#1} \left (#2\right)}}

\newcommand{\srcrossent}[3][r]{\ensuremath{\tilde X_{#1} \left (#2\|#3\right)}}
\newcommand{\srdiv}[3][r]{\ensuremath{\tilde D_{#1}\left(#2\|#3 \right)}}

\newcommand{\srdist}[2][r]{\ensuremath{\tilde q_{#1}(#2)}} 

\Title{The R\'enyi Entropies operate in 
Positive Semifields%
}

\Author{Francisco J.~Valverde-Albacete$^{1,\ddagger}$\orcidA{}, Carmen~Pel\'aez-Moreno$^{2,\ddagger}$*\orcidB{}}

\AuthorNames{Francisco J.~Valverde-Albacete and Carmen~Pel\'aez-Moreno}

\address{%
$^{1}$ \quad Department of Signal Theory and Communications, Universidad Carlos III de Madrid, Legan\'es 28911, Spain; fva@tsc.uc3m.es\\
$^{2}$ \quad Department of Signal Theory and Communications, Universidad Carlos III de Madrid, Legan\'es 28911, Spain; carmen@tsc.uc3m.es}

\corres{Correspondence: carmen@tsc.uc3m.es; Tel.: +34-91-624-8771}

\firstnote{These authors contributed equally to this work.}

\abstract{%
We set out to demonstrate that the R\'enyi entropies with shifted parameter $r=\alpha -1$ are better thought of as operating in a type of non-linear semiring called a positive semifield.  
We show how the R\'enyi's postulates lead to Pap's g-calculus where the functions carrying out the domain transformation are R\'enyi's information function and its inverse. In its turn, Pap's g-calculus under R\'enyi's information function transforms the set of positive reals into a family of semirings where ``standard'' product has been transformed into sum and ``standard''  sum into a power-emphasized  sum. Consequently, the transformed product has an inverse whence the structure is actually that of a positive semifield. 
Instances of this construction lead to idempotent analysis and tropical algebra as well as to less exotic structures. 
We conjecture that this is one of the reasons why tropical algebra procedures, like the Viterbi algorithm of dynamic programming,  morphological processing, or neural networks are so successful in computational intelligence applications. 
But also, why there seem to exist so many procedures to deal with ``information'' at large. 
}


\begin{document}

\section{Introduction}
\label{sec:intro}
Some non-linear algebras like the max-plus semiring or, in general, positive semirings have wide application in Artificial Intelligence, Machine Learning, and Computational Intelligence. 
In this paper we propose the following explanation for their ubiquity: that they are in fact the natural algebras in which  each of the infinite instances of the R\'enyi informations operate so that applications are ``matched'' to specific values of the order parameter. 
From this starting point, the quasi-genetic social process of scientific and technological advance would then select which value for $\alpha$ is most suited to make a technique work in a particular application. 
We will argue in this paper that these non-standard algebras are the positive semifields, a special type of semiring with multiplicative inverses but no additive inverses. 

Recall that a  \emph{semiring} is an algebra  $\mathcal S = \langle S,\oplus,\otimes,\epsilon,e\rangle$ whose
 additive structure $\langle S, \oplus, \epsilon \rangle$ is a commutative monoid and whose multiplicative structure $\langle S\backslash\{\epsilon\}, \otimes, e\rangle$ is a monoid with multiplication distributing over addition from right and left 
and an additive neutral element 
absorbing for $\otimes$, i.e. $\forall a \in S,\; \epsilon \otimes a = \epsilon$~\cite{gol:99a}. 
A semiring is \emph{commutative} if its product is commutative. All semirings considered in this paper are commutative, whence we will drop the qualification altogether. 
A semiring is \emph{zero sum-free} if whenever a sum is null all summands are also null, 
and \emph{entire} if it has no non-null factors of zero.
A semiring is \emph{positive} if it is zero sum-free and entire~\cite{gon:min:08}. 
%
%
Finally, a semiring is a \emph{semifield} if there exists a
multiplicative inverse for every element $a \in S$---notated as $\linv
a$--- except the null. 
All semifields are {entire}.

In this paper we concern ourselves with  \emph{positive semifields}, e.g. entire, zero sum-free semirings with a multiplicative inverse but no additive inverse (see \S~\ref{sec:pos:smf}), whose paragon is the set of non-negative reals with its standard operations $\nnR = \langle [0,\infty), +, \times, \linv{\cdot}, 0, 1 \rangle$. 
%
Their interest lies in the fact that positive semifield applications abound in many areas of research. 
To cite but a few examples:
\begin{itemize}
\item 
Artificial Intelligence (AI)~\cite{rus:nor:10} is an extensive field under which applications abound dealing with minimizing costs or maximizing utilities. Semifields and their dual-orderings (see Sections~\ref{sec:pos:smf} and~\ref{sec:construction}) provide a perspective to mix these two kinds of valuations. 

\item 
%
Machine learning (ML)~\cite{mur:12} makes heavy use of Probability Theory, which is built around the positive semifield of the non-negative reals with their standard algebra $\nnR$ and negative logarithms thereof---called \emph{log-probabilities or log-likelihoods} depending on the point of view---both of which are positive semifields, as shown in Sections~\ref{sec:pos:smf} and~\ref{sec:basic:entsmf}. 

\item 
%
Computational Intelligence (CI)~\cite{eng:02} makes heavy use of positive semirings in the guise of \emph{fuzzy semirings}. Although semifields cannot be considered ``fuzzy'' for several technical reasons, the name is sometimes an umbrella term under which  \emph{non-standard algebras} are included, many of which are semifields, e.g. the morphological semifield of morphological processing and memories, a special case of the semifields in Section~\ref{sec:entropy:smf}. 

\item Other applications of positive semifields not related to modeling intelligence include Electrical Network analysis and synthesis (see the example in Section~\ref{sec:pos:smf}), 
queuing theory~\cite{BCOQ92} and flow shop scheduling~\cite{but:10}. 

\end{itemize}

To build a basis for our initial conjecture,  
we first revisit  a couple of seemingly unrelated topics: 
first, a shifting $r = \alpha -1$  of the index parameter of the  R\'enyi $\alpha$-entropies, our model of information measure for the present purposes:
\begin{align}
\label{eq:RenyE}
&
H_\alpha(P_X) = \frac{1}{1-\alpha}\log\left(\sum_{i=1}^np_i^\alpha\right)\,.
\end{align}
that makes the connection between these entropies and the weighted H\"older means transparent. 
And second, Pap's g-calculus as a construction on positive semifields. 

In our results we briefly
introduce an ``entropic'' semifield 
prior to proving  how the shifted R\'enyi entropy takes values in positive semifields, 
and also how this reflects back on the semifield of positive numbers interpreted as ``probabilities''. 
This allows us to unfold the argumentation for our conjecture that AI, ML and CI applications are mostly dealing with R\'enyi entropies of different order. 
We end the paper with a discussion of the issues touched upon it and some conclusions. 

%
%
\section{Materials and Methods}
\label{sec:methods}
%

\subsection{The Shifted R\'enyi entropy}
\label{sec:renyi:entropy}
%
%

Recall that the \emph{weighted power or H\"older mean of order $r$}~\cite{har:lit:pol:52} is defined as 
\begin{align}
\label{eq:holder:wmean}
\hmean{\vec w}{\vec x} = \left( \frac{\sum_{i=1}^n w_i\cdot x_i^r}{\sum_k w_k}\right)^\frac{1}{r}
= \left( \sum_{i=1}^n \frac{w_i}{\sum_k w_k}\cdot x_i^r\right)^\frac{1}{r}
\end{align}
When $r\rightarrow 0$ the geometric mean appears:
\[
\hmean[0]{\vec w}{\vec x} = \lim_{r\rightarrow 0} \hmean{\vec w}{\vec x} = \left(\Pi_{i=1}^n x_i^{w_i}\right)^\frac{1}{\sum_k w_k}
\]

To leverage the theory of generalized means to our advantage, 
we start with a correction to R\'enyi's entropy definition: in~\cite{val:pel:19a} a case is made for shifting the original statement  from the index that R\'enyi proposed $\alpha$  to $r = \alpha - 1$. 
Specifically, the transformation function for the average of surprisals~\cite{ren:70} in the  R\'enyi entropy~\eqref{eq:RenyE} is arbitrary in the parameter $\alpha$ choosen for it and 
we may substitute $r = \alpha - 1$ to obtain the pair of formulas:
\begin{align}
\label{eq:adapting}
\varphi'(h) &= b^{-rh}
&
\varphi'^{-1}(p) &= \frac{-1}{r}\log_b p  
\end{align}
Note that the basis of the logaritm is unimportant. As customary in Information Theory, we presuppose $\log x \equiv \log_2 x$.
%
 The following definitions are thus obtained: 
\begin{Definition}
\label{def:sRenyie}
Let  $P_X(x_i) = p_i$ and $Q_X(y_i) = q_i$ be two distributions with compatible support. Then 
the expression of the 
\emph{
shifted R\'enyi 
entropy $\srentropy{P_X}$, 
cross-entropy $\srcrossent{P_X}{Q_X}$, 
and divergence $\srdiv{P_X}{Q_X}$} are: 
\begin{align}
\label{eq:sRenyE:entropies}
\srentropy{P_X} &= - \log {\hmean{P_X}{P_X}} 
\\
\label{eq:sRenyC:entropies}
\srcrossent{P_X}{Q_X}  &= - \log \hmean{P_X}{Q_X}
\\
\label{eq:sRenyD:entropies}
\srdiv{P_X}{Q_X} & = \log \hmean{P_X}{\frac{P_X}{Q_X}}
\end{align}
\end{Definition}

Several brief points are worth mentioning in this respect, although the full argument can be followed in~\cite{val:pel:19a}:
\begin{itemize}
\item 
Important cases of the means for historical and practical reasons and their relation to the (shifted and original) R\'enyi entropy are shown in Table~\ref{tab:entropies} and their relationship to the entropies. 
\begin{table*}
\centering
\small
\begin{tabular}{|l|c|l|l|r|r|}
\hline
Mean name & Mean $\hmean[r]{\vec w}{\vec x}$& Shifted entropy $\pentropy[r]{P_X}$ & Entropy name & $\alpha$ & $r$\\
\hline
Maximum & $\max_i x_i$ & $\tilde H_{\infty} = -\log \max_i p_i$ & min-entropy & $\infty$ & $\infty$ \\
\hline
Arithmetic & $ \sum_i w_i x_i $ & $\tilde H_{1} =-\log \sum_i p_i^2$ & R\'enyi's quadratic & $2$ & $1$ \\
\hline
Geometric & $\Pi_i x_i^{w_i}$ & $\tilde H_{0} = -\sum_i p_i \log p_i$ & Shannon's & $1$ & $0$\\
\hline
Harmonic & $ (\sum_i w_i \frac{1}{x_i})^{-1}$ & $\tilde H_{-1} = \log n$  & Hartley's & $0$ & $-1$\\
\hline 
Minimum & $\min_i x_i$ & $\tilde H_{-\infty} = -\log \min_i p_i$ & max-entropy & $-\infty$ & $-\infty$\\
\hline
\end{tabular}
\caption{Relation between the most usual weighted power means, R\'enyi entropies and shifted versions of them, from~\protect{\cite{val:pel:19a}}.}
\label{tab:entropies}
\end{table*}

\item The properties of the R\'enyi entropy, therefore, stem from those of the mean, inversion and the logarithm%
\iftoggle{extendedVersion}{:%

\begin{Proposition}[Properties of the R\'enyi spectrum of $P_X$]
\label{prop:prop:sE}
Let $r,s \in \mathbb R\cup\{\pm\infty\}$, and 
$P_X, Q_X \in \Delta^{n-1}$ where $\Delta^{n-1}$ 
is the simplex over the support $\supp{X}$, with cardinal $|\supp{X}| = n$. Then, 
\begin{enumerate}
\item \label{prop:prop:sE:p11} (Monotonicity) 
	The R\'enyi entropy is a non-increasing function of the order $r$.
\begin{align}
	s \leq r \Rightarrow \srentropy[s]{P_X} \geq \srentropy[r]{P_X} 
\end{align}

\item \label{prop:prop:sE:p12} (Boundedness) 
The R\'enyi spectrum $\srentropy{P_X}$ is bounded by the limits
\begin{align}
	\srentropy[-\infty]{P_X}  \geq \srentropy[r]{P_X} \geq \srentropy[\infty]{P_X} 
\end{align}

\item \label{prop:prop:sE:p2} 
	The entropy of the uniform \emph{pmf} $U_X$ is constant over $r$\,.
\begin{align}
	\forall r \in \mathbb R\cup\{\pm\infty\}\,,  \srentropy[r]{U_X} = \log n
\end{align}

\item \label{prop:prop:sE:p3} 
	The Hartley entropy ($r=-1$) is constant over the distribution simplex. 
\begin{align}	
	\srentropy[-1]{P_X} = \log n 
\end{align}
\item \label{prop:prop:sE:p5} (Divergence from uniformity) 
The divergence of any distribution $P_X$ from the uniform $U_X$ can be written in terms of the entropies as: 
\begin{align}
	\srdiv[r]{P_X}{U_X} = \srentropy{U_X} - \srentropy{P_X}\,.
\end{align}

\item \label{prop:prop:sE:p7} (Derivative of the shifted entropy) 
The derivative in $r$ of R\'enyi's $r$-th order entropy is
\begin{align}
\frac{d}{dr}\srentropy{P_X} = \frac{-1}{r^2}\srdiv[0]{\srdist[r]{P_X}}{P_X}
	= \frac{-1}{r}\log\frac{\hmean[0]{\srdist[r]{P_X}}{P_X}}{\hmean{P_X}{P_X}}\,,
\end{align}
where $\srdist[r]{P_X}=\left\{\frac{p_ip_i^r}{\sum_k p_k p_k^r}\right\}_{i=1}^n$\,.

\item \label{prop:prop:sE:p6}  (Relationship with the moments of $P_X$) 
The shifted R\'enyi Entropy of order $r$ is the logarithm of the inverse $r$-th root of the $r$-th moment of $P_X$\,.
\begin{align}
\label{eq:sre:pot}
\srentropy{P_X} &= -\frac{1}{r} \log E_{P_X}\{P_X^r\} = \log \frac{1}{\sqrt[r]{ E_{P_X}\{P_X^r\} }}
\end{align}

\end{enumerate}
\end{Proposition}
where we have introduced the notion of \emph{shifted escort probabilities} $\srdist[r]{P_X}$ acting in the shifted R\'enyi entropies as the analogues of the \emph{escort probabilities} in the standard definition~\cite{bec:sch:95} 
$(\srdist{P_X})_i = \frac{p_i p_i^r}{\sum_k p_k p_k^r} = \frac{p_i^{\alpha}}{\sum_k p_k^{\alpha}}=(\rdist{P_X})_i$.  
 They are  just the shifting of the traditional escort probabilities. 
}{.}

\item This is not merely a cosmetic change, since it has the potential to allow the simplification of issues and the discovery of new ones in dealing with the R\'enyi magnitudes. For instance,
since the means are defined for all $r \in [-\infty, \infty]$ there cannot be any objection to considering negative values for the index of the shifted entropy. This motivates calling $\srentropy{P_X}$ the \emph{R\'enyi spectrum (of entropy).}

\item The definition makes it also evident that the shifted cross-entropy seems to be the more general concept, given that the shifted entropy and divergence are clearly instances of it.
\begin{align*}
\srentropy{P_X} &= \srcrossent{P_X}{P_X}
&
\srdiv{P_X}{Q_X}  & = - \srcrossent{P_X}{P_X/Q_X} 
\end{align*}

\item Also, the following lemma from~\cite{val:pel:19a} shows that the R\'enyi entropies can be rewritten in terms of the Shannon cross entropy and the Kullback-Leibler divergence:
\begin{Lemma}
\label{lemma:ren:shan}
Let $r,s \in \mathbb R\cup\{\pm\infty\}$, 
$P_X \in \Delta^{n-1}$ where $\Delta^{n-1}$ is the simplex over the support $\supp{X}$%
\iftoggle{extendedVersion}{%
}{%
 and $\srdist[r]{P_X}=\left\{\frac{p_ip_i^r}{\sum_k p_k p_k^r}\right\}_{i=1}^n$
 are the \emph{escort probabilities}~\protect{\cite{val:pel:19a,bec:sch:95}}
}
. Then, 
\begin{align}
\label{eq:ren:shan:div}
\srentropy{P_X} &= \frac{1}{r}\srdiv[0]{\srdist{P_X}}{P_X}  + \srcrossent[0]{\srdist{P_X}}{P_X}
\\
\label{eq:ren:shan:ent}
\srentropy{P_X} &= \frac{-1}{r}\srentropy[0]{\srdist{P_X}}  + \frac{r+1}{r}\srcrossent[0]{\srdist{P_X}}{P_X}
\end{align}
\end{Lemma}
\noindent
 
 \item Lemma~\eqref{lemma:ren:shan} rewrites the entropies  in terms of the geometric means which is, by no means, the only rewriting possible. Indeed, some would say that the arithmetic mean is more natural, and this is the program of Information Theoretic Learning~\cite{pri:10}, where it is explored under the guise of $\rentropy[2]{P_X} \equiv \srentropy[1]{P_X} $. 

\item The shifting clarifies the relationship between quantities around the R\'enyi entropy. 
For instance,  due to Hartley's function---the surprisal function---from every average measure of information, an equivalent \emph{average} probability emerges: define the extension to Hartley's information function to non-negative numbers $\hinf{\cdot}: [0,\infty] \rightarrow [-\infty, \infty]$ as $\hinf{p}= - \ln p$. This is  one-to-one from $[0,\infty]$ and total onto $[-\infty, \infty]$, with inverse $\hinfinv{h} = e^{-h}$ for $h \in [-\infty, \infty]$.  
\begin{Definition}
Let $X \sim P_X$ with R\'enyi spectrum $\srentropy[r]{P_X}$. 
Then the \emph{equivalent probability function of $\srprob[r]{P_X}$} is the Hartley inverse of $\srentropy[r]{P_X}$ over all values of $r \in [-\infty, \infty]$ 
\begin{align}
\label{eq:eq:prob}
\srprob[r]{P_X} = \hinfinv{\srentropy[r]{P_X}} = \hmean{P_X}{P_X}
\end{align}
\end{Definition}

\item Similarly, the information potential $\srpot{P_X} = E_{P_X}\{P_X^r\}$ has been independently motivated in a number of applications~\cite{pri:10}.
The next lemma is immediate using the conversion function \eqref{eq:adapting} on the shifted entropy~\eqref{eq:sRenyE:entropies}. 

\begin{Lemma}
\label{lemma:part:pot}
Let $X \sim P_X$. The \emph{information potential} is the $\varphi'$ image  of the shifted R\'enyi entropy 
\begin{align}
\srpot{P_X}  = E_{P_X}\{P_X^r\} = \sum_i \frac{p_i}{\sum_k p_k} p_i^r = b^{-r\srentropy{P_X}} = \varphi'(\srentropy{P_X})
\end{align}
\end{Lemma}

\end{itemize}

\begin{figure}
\begin{subfigure}[b]{0.5\textwidth}
	\centering
	\begin{tikzpicture}[scale=1.0]
	\node (P)  at (3,2) {$\srprob[r]{P_X}$};
	\node (H) at (0,0) {$\srentropy{P_X }$};
	\node (V)  at (3,-2) {$\srpot{P_X}$};
	\path[->,font=\scriptsize,>=angle 90]
	(P) edge  [bend left=10] node[below right]{$\mathfrak I^\ast$} (H)
	(H) edge  [bend left=10] node[above right]{$\varphi'$} (V)
	(V) edge  [bend left=10] node[left]{$\cdot^{r}$} (P)
	(P) edge [bend left=10] node[right]{$\cdot^{1/r}$} (V)
	(V) edge [bend left=10] node[below left]{$\varphi'^{-1}$} (H)
	(H) edge [bend left=10] node[above left ]{${\left(\mathfrak I^\ast\right)}^{-1}$} (P)
	;
	\end{tikzpicture}
	\caption{Between shifted entropy-related quantities}
	\label{fig:circaREntropy:magnitudes}
\end{subfigure}
\begin{subfigure}[b]{0.5\textwidth}
	\centering
	\begin{tikzpicture}[scale=1.0]
	\node (P)  at (3,2) {$\nnR$};
	\node (H) at (0,0) {$\mathbb H$};
	\node (P_r)  at (3,-2) {$\nnR^r$};
	\path[->,font=\scriptsize,>=angle 90]
	(P) edge  [bend left=10] node[below right]{$\mathfrak I^\ast$} (H)
	(H) edge  [bend left=10] node[above right]{$\varphi'$} (P_r)
	(P_r) edge  [bend left=10] node[left]{$\cdot^{r}$} (P)
	(P) edge [bend left=10] node[right]{$\cdot^{1/r}$} (P_r)
	(P_r) edge [bend left=10] node[below left]{$\varphi'^{-1}$} (H)
	(H) edge [bend left=10] node[above left ]{${\left(\mathfrak I^\ast\right)}^{-1}$} (P)
	;
	\end{tikzpicture}
	\caption{Between entropy-related domains (see \S~\ref{sec:entropy:smf})}
	\label{fig:circaREntropy:domains}
\end{subfigure}

\caption[]{Schematics of relationship due to entropic isomorphisms}
\label{fig:circaREntropy}
\end{figure}
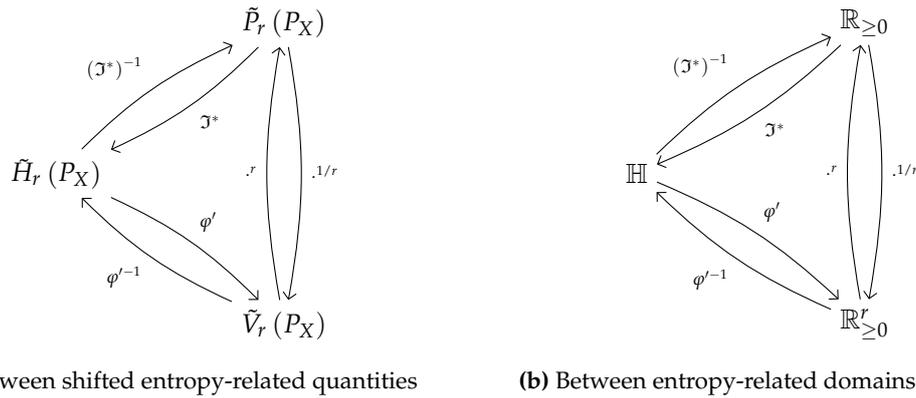

These three quantities---the shifted R\'enyi entropy, the equivalent probability function, and the information potential---%
stand in a relationship, as described in Figure~\ref{fig:circaREntropy}.\subref{fig:circaREntropy:magnitudes}%
\footnote{A similar diagram is, of course, available for the standard entropy, using $\varphi$ with the $\alpha$ parameter. },  whose characterization is the conducting thread of this paper.
In~\cite{val:pel:19a} other formulas to convert them into each other are tabulated.  

%
%
%
%

\subsection{Positive Semifields}
\label{sec:pos:smf}
From the material in Section~\ref{sec:renyi:entropy} it seems evident that non-negative quantities are important for our purposes. 
Non-negativity is captured by  the concept of \emph{zero sum-free semiring} mentioned above, but we focus in the slightly less general notion of \emph{dioid} (for double monoid)~\cite{gon:min:08} where there is an order available that interacts ``nicely'' with the operations of the algebra.  
\label{sec:smr:smf}
\subsubsection{Complete and positive dioids}
A  \emph{dioid} is a commutative semiring $\mathcal D$ where  the canonical preorder relation---$a \preccurlyeq b$ if and only if there exists $c \in D$ with $a \oplus c = b$---is actually an order $\langle D, \preccurlyeq\rangle$\,. 
In a dioid the canonical orden relation is compatible with both $\oplus$ and $\otimes$~\cite[Chap. 1, Prop. 6.1.7]{gon:min:08} 
and the additive zero $epsilon$ is always the infimum of the dioid or \emph{bottom} hence the notation $\epsilon=\inf D=\bot$\,.
Dioids are
all zero sum-free, that is, they have no non-null additive factors of zero: if $a,b \in D, a \oplus b = \epsilon$ then $a = \epsilon$ \emph{and} $b = \epsilon$\,. 
%

A dioid is \emph{complete} if it is complete as an ordered set for the canonical order relation, and the following distributivity properties hold, for all $A \subseteq D, b \in D$,
\begin{align}
\left( \bigoplus_{a \in A} a\right) \otimes b &= \bigoplus_{a \in A} (a \otimes b)
&
b \otimes \left( \bigoplus_{a \in A} a\right)&= \bigoplus_{a \in A} (b \otimes a)
\end{align}
In {complete} dioids, there is already a top element $\top=\oplus_{a \in D} a$\,. 


A semiring is \emph{entire or zero-divisor free} if $a \otimes b = \epsilon$ implies $a = \epsilon$ or $b = \epsilon$\,. 
If the dioid is entire, its order properties justifies calling it a \textit{positive dioid} or \textit{information algebra}~\cite{gon:min:08}. 

\subsubsection{Positive semifields}
A semifield, as mentioned in the introduction, is a semiring whose multiplicative structure $\langle K\setminus\{\epsilon\}, \otimes, e, \lconj{\cdot}\rangle$ is a group, where $\lconj{\cdot}: K \rightarrow K$ is the function to calculate the inverse such that $\forall u \in K, u \otimes \lconj{u} = e$\,. 
Since all semifields are entire, dioids that are at the same time semifields are called \emph{positive semifields}, of which the positive reals or rationals are a paragon.

\begin{Example}[Semifield of non-negative reals]
\label{exa:nnR}
The nonnegative reals
$\nnR=\langle [0,\infty), +, \times , \lconj{\cdot}, \bot=0, e=1 \rangle$
are the basis for the computations in Probability Theory and other quantities that are multiplicatively aggregated. The $0$ has no inverse hence $\nnR$ is incomplete. 
Consider modeling \emph{utilities} and \emph{costs} with this algebra. 
For utilities $0$ acts as a least element---a \emph{bottom}---and the order is somewhat directed ``away'' from this element hence the underlying order is $\langle [0, \infty), \leq \rangle$, that is to say \emph{utilities are to be maximized}. 
But its dual order $\langle [0, \infty), \geq \rangle$---intended to model \emph{costs}, \emph{to be minimized}---has no bottom (see below), hence applications using, for instance, multiplicative costs and utilities at the same time, will be difficult to carry out in this algebra and notation. 
\qed
\end{Example}

\begin{figure}[!ht]
  \centering
 \includegraphics[scale=0.76]{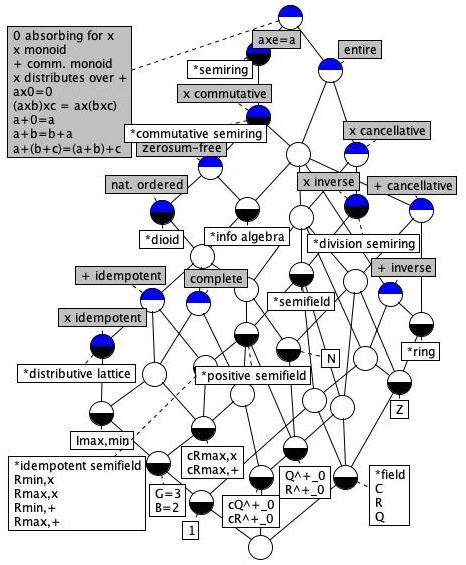}
  \caption[]{\textbf{Lattice of a selection of abstract (leading asterisk, white label) and concrete (white label) commutative semirings and their properties (grey label) mentioned in the text. Adapted from~\cite{val:pel:15a}.} %
Each node is a concept of Abstract Algebra: its properties are obtained from the gray labels in nodes upwards, and its structures from the white labels in nodes downwards. 
The picture is related to the \emph{chosen} sets of properties and algebras and does not fully reflect the structure of the class of semirings. 
We have chosen to highlight complete semifields, like $\nnR$.
}%
  \label{fig:comm_pos_smfs}%
\end{figure}

Figure~\ref{fig:comm_pos_smfs} shows a concept lattice of the position of positive semifields within the commutative semirings, as well as the better known collateral families of fields like $\mathbb R$ and distributive lattices like 
$\langle [0,1], \max, \min\rangle$
(cfr.~\cite[Fig.~1]{val:pel:15a}).



In incomplete semifields like the one above, the inverse of the bottom element is the ``elephant in the room'' to be avoided in computations. Fortunately, semiring theory provides a construction to supply this missing element in semifields~\cite{gol:99a}. 
However, the problem with the dual order in the semifield mentioned above suggests that we introduce \emph{both} completions at the same time through  the following theorem\footnote{The need for the dotted notation will be made clear after the theorem.}. 
\begin{Theorem}
\label{theo:order:duals}
For every incomplete positive semifield $\mathcal K = \langle K, \oplus, \otimes, \lconj{\cdot}, \bot, e\rangle$ 
\begin{enumerate}

\item \label{theo:dualsmf:1}
There is a pair of completed semifields over $\overline K = K \cup \{\top\} $
\begin{align}
\cK &= \langle K, \oplusl, \otimesl, \lconj{\cdot}, \bot, e, \top\rangle
&
\lconj{\cK} &= \langle K, \oplusu, \otimesu, \lconj{\cdot}, \top, e, \bot\rangle
\end{align}
where $\top = \lconj \bot$ and $\bot = \lconj \top$ by definition,  

\item \label{theo:dualsmf:3}
In addition to the individual laws as positive semifields, we have the modular laws:
\begin{align}
\label{eq:dualsmf:modular}
(u \oplusl v) \otimesl (u \oplusu v) &= u \otimesl v
&
(u \oplusl v) \otimesu (u \oplusu v) &= u \otimesu v
\intertext{the analogues of the De Morgan laws:}
\label{eq:dualsmf:DeMorgan}
u \oplusl v &= \lconj{(\lconj{u} \oplusu \lconj{v})}
&
u \oplusu v &= \lconj{(\lconj{u} \oplusl \lconj{v})}
\\
u \otimesl v &= \lconj{(\lconj{u} \otimesu \lconj{v})}
&
u \otimesu v &= \lconj{(\lconj{u} \otimesl \lconj{v})} \notag
\intertext{and the self-dual inequality in the natural order}
\label{eq:dualsmf:inequality}
\centering
u \otimesl (v \otimesu w)
&\preccurlyeq 
(u \otimesl v) \otimesu w \,.
\end{align}

\item \label{theo:dualsmf:2}
Further, if $\mathcal K$ is a positive dioid, then 
the inversion operation is a dual order isomorphism between the dual order structures $\cK = \langle K, \preccurlyeq \rangle$ and $ \lconj{(\cK)} = \langle K, \succcurlyeq \equiv \preccurlyeq^\delta \rangle$ 
with the natural order of the original semifield a suborder of the first structure.

\end{enumerate}
\end{Theorem}
%
%
\begin{proof}
For~\ref{theo:dualsmf:1}, 
%
%
consider $\overline K = K \cup \{\top\}$ obtained by adding a top element to $K$. 
The order $\langle \overline K, \preccurlyeq \rangle$ is extended with  $x \preccurlyeq \top, \forall x \in K$, hence the notation $\top$ for this element. 
%
%
In this completed set the definition of the operations are\footnote{In this paper, cases in a definition-by-case should be interpreted from top to bottom: this first case to match applies.}:
\begin{align*}
u \oplusl v &=
\begin{cases}
	\top & u = \top \text{ or } v = \top\\
	u \oplus v & u,v \in K\setminus\{\top\}\\
\end{cases}
&
u \otimesl v &=
\begin{cases}
	\bot &	u = \bot \text{ or } v = \bot\\
	\top & u = \top \text{ or } v = \top\\
	u \otimes v & u,v \in K\setminus\{\bot,\top\}\
\end{cases}
\end{align*}
with the inversion operation completed by the definition of $\top = \lconj \bot$ and $\bot = \lconj \top$.
So $\cK = \langle K, \oplusl, \otimesl, \lconj{\cdot}, \bot, e, \top\rangle$ is the well-known top-completion of a positive semiring~\cite[p.~250]{gol:99a}. 

In this construction, inversion is total, injective and surjective in the completed domain $\overline K$, hence a bijection. It is easily seen as an involution $\lconj{(\lconj x)} = x, \forall x \in \overline K$, in fact, the inverse for the order-dual $\lconj{\overline{\mathcal K}}$\,. 
But, the operations in the inverse semifield are given by:
\begin{align}
a \oplusu b &= \lconj{\left(\lconj a \oplusl \lconj b\right)}
&
a \otimesu b &= \lconj{\left(\lconj a \otimesl \lconj b\right)}
\end{align}
To gain an understanding of the operations, we explore their results on a case-by-case basis. For $\oplusu$:
\begin{itemize}
\item If $a = \top$ then $\lconj a \oplusl \lconj b = \bot \oplusl \lconj b = \lconj b$, whence $\top \oplusu b = b$, and symmetrically for $b= \top$. That is, $\top$ is the neutral element of addition $\oplusu$ and $\langle \overline K, \oplusu, \top \rangle$ a monoid. 

\item If $a = \bot$, then $\lconj \bot \oplusl \lconj b = \top \oplusl \lconj b = \top$, whence $\bot \oplusu b = \lconj \top = \bot$ and symmetrically for $b = \bot$\,.
This proves that $\bot$ is the maximum element of $\preccurlyeq^\delta$, to be defined below. 

\item Otherwise, for  $\{a,b\} \subseteq K\setminus\{\bot\}$, we have $a \oplusu b = \frac{1}{\lconj a \oplus \lconj b} = \frac{a \otimes b}{a \oplus b}$\,,
 
\end{itemize}
while for $\otimesu$:
\begin{itemize}
\item If $a = \top$ then $\lconj a \otimesl \lconj b = \bot \otimesl \lconj b = \bot$,  whence $\top \otimesu b = \lconj \bot = \top$, and symmetrically for $b$.

\item If $a = \bot$ but $b \neq \top$, then $\lconj a \otimesl \lconj b = \top \otimesl \lconj b = \top$, whence $a \otimesu b = \bot$. 

\item Otherwise, for  $\{a,b\} \subseteq K\setminus\{\bot\}$, we have $a \otimesu b = a \otimes b = a \otimesl b$.

\end{itemize} 
Note that $\langle \overline K, \oplusu, \top\rangle$ is a commutative monoid, with commutativity and associativity following from those of $\oplusl$. 
Likewise,  $\lconj e = e$ is easily proven to be the neutral element of $\langle \overline K, \otimesu, e, \lconj{\cdot}\rangle$ which is a commutative group with commutativity and associativity issuing from those of $\otimesl$, whose inverse is the involution $\lconj{\cdot}$. So to prove that the algebra is a semifield we only need to prove the distributive of $\otimesu$ over $\oplusu$, for $u,v,z \in \overline K$: 
\begin{align*}
u \otimesu (v \oplusu z) 
	&= \frac{1}{\lconj u \otimesl \lconj{\left(\lconj{\left( \lconj v \oplusl \lconj z\right)}\right)}}
	=  \frac{1}{\lconj u \otimesl (\lconj v \oplusl \lconj z)}	
	= \frac{1}{\lconj u \otimesl \lconj v \oplusl \lconj u \otimesl \lconj z} 
	\\
	&= \lconj{(\lconj u \otimesl \lconj v)} \oplusu \lconj{(\lconj u \otimesl \lconj z)}
	= u \otimesu v \oplusu u \otimesu z
\end{align*} 
Therefore $\lconj{\cK} = \langle K, \oplusu, \otimesu, \lconj{\cdot}, \top, e, \bot\rangle$ is another completed semifield issuing from the first one. 

For~\ref{theo:dualsmf:3}, the proof for the De Morgan-like laws is easy from the definion of $\oplusu$ and $\otimesu$: those definitions are actually one half of the laws, e.g. 
$a \otimesu b = \lconj{\left(\lconj a \otimesl \lconj b\right)}$.
Inverting and by the involutivity of the inversion $\lconj a \otimesl \lconj b = \lconj{( a \otimesu b)}$, so we change $a = \lconj u$ and $b = \lconj v$ to prove the result. 
The proof is analogue for the multiplicative law. 
And the dual equalities~\eqref{eq:dualsmf:modular} and the self-dual inequality~\eqref{eq:dualsmf:inequality} are just exercises in case analysis. 

For~\ref{theo:dualsmf:2} we want to find the order for the dual semifield, $\dual\preccurlyeq = \succcurlyeq$.
\begin{itemize}
\item  if $\{u,v\} \subseteq K \setminus\{\bot\}$ , since the natural order is compatible with multiplication we multiply by $\lconj u\otimesl \lconj v$ to obtain $u \otimes (\lconj u \otimes \lconj v) \preccurlyeq v \otimes (\lconj u \otimes \lconj v)$
whence, by cancellation, $\lconj v \preccurlyeq \lconj u$, or else $\lconj u \dual{\preccurlyeq} \lconj v$, so the order is the dual on inverses. 

\item We have that $\bot \preccurlyeq v, \forall v \in \overline K$, otherwise $v \dual{\preccurlyeq} \bot$ which asserts that $\bot = \lconj \top$ is the ``top'' of the inverted order. 
Likewise we read from $u \preccurlyeq \top, \forall u \in \overline K$ that $\top \dual{\preccurlyeq} u$, that is $\top=\lconj \bot$ is the ``bottom'' in $\dual{\preccurlyeq}$. 
\end{itemize}
whence $\lconj{\langle \overline K, \preccurlyeq \rangle} = \langle \overline K, \dual{\preccurlyeq}  \rangle$. 
\end{proof}
%
%
This proof provides extensive guide on how to use the notation. Note that:
\begin{itemize}
\item The dot notation, from~\cite{mor:70}, is a mnemonic for where do the multiplication of the bottom and top go:
\begin{align*}
\bot \otimesl \top &= \bot
&
\bot \otimesu \top &= \top
\end{align*}
implying that the ``lower'' addition and multiplication are aligned with the original order in the semifield, e.g. $\bot \otimesl x = \bot$  while the ``upper'' addition and multiplication are aligned with its dual. All other cases remain as defined for $\oplus$ in the incomplete semifield. 

\item The case analysis for the operators in the dual semifield allows us to write their definition-by-cases as follows:
\begin{align}
\label{eq:dual:operations}
a \oplusu b &=
\begin{cases}
b & a = \top\\
a & b = \top\\
\bot & a = \bot \text{ or } b = \bot\\
\frac{1}{\lconj a \oplus \lconj b} 
& \{a,b\} \subseteq K\setminus\{\bot\}\
\end{cases}
&
a \otimesu b  &=
\begin{cases}
	\top & a = \top \text{ or } b = \top\\
	\bot &	a = \bot \text{ or } b = \bot\\
	a \otimes b &  \{a,b\} \subseteq K\setminus\{\bot\}\
\end{cases}
\end{align}
This is important for calculations, 
but notice that $\otimesu$ and $\otimesl$ only differ in the corner cases.

\item The notation to ``speak'' about these semirings tries to follow a convention reminiscent of that of boolean algebra, where the inversion is the complement~\cite[Ch.~12]{elle:95}.

\item Note that $\oplusl$ and $\oplusu$ seem to operate on different ``polarities'' of the underlying set: if one operates on two numbers, the other operates on their inverses while this is not so for the respective multiplications.  
This proves extremely important to model physical quantities and other concepts with these calculi (see example below).
\end{itemize} 

Regarding the intrinsic usefulness of completed positive semifields that are not fields---apart from the very obvious but degenerate case of $\mathbb B$, the booleans---we have the following example used, for instance, in Convex Analysis and Electrical Network theory. 
\begin{Example}[Dual semifields for the non-negative reals]
\label{exa:dual:smflds:nnR}
The previous procedure shows that there are some problems with the notation of Example~\ref{exa:nnR}, and this led to the definition of the following signatures for this semifield and its inverse in convex analysis~\cite{mor:70}:
\begin{align}
\label{eq:pair:nnR}
\nnR &= \langle [0,\infty], \plusl, \timesl, \linv{\cdot}, 0, 1, \infty\rangle
&
\linv{\nnR} &= \langle [0,\infty], \plusu, \timesu, \linv{\cdot}, \infty, 1, 0\rangle
\end{align}

\noindent
Both of these algebras are used, for instance, in  Electrical Engineering (EE), 
the algebra of complete positive reals to carry out the series summation of resistances, and its dual semifield to carry out parallel summation of conductances.
With the convention that $\nnR$ semiring models \emph{resistances}, it is easy to see that the bottom element, $\bot = 0$ models a shortcircuit, that the top element $\top = \infty$ models an open circuit (infinite resistance) and these conventions are swapped in the dually-ordered semifield of {conductances}.
Since EE does not use the dotted notation explained in this paper, 
the formulas required for the multiplication of the extremes:
\begin{align*}
0 \timesl \infty &= 0
&
0 \timesu \infty &= \infty
\end{align*}
aren't allowed in 
circuit analysis.
In our opinion, this strongly suggests that what is actually being operated with are the incomplete versions of these semifields, and the many problems that EE students have in learning how to properly deal with these values may stem from this fact. 
Other uses in Economics are detailed in~\cite{elle:95}.
\qed
\end{Example}

\begin{Example}[Multiplicatively and additively idempotent costs and utilities]
Several pairs of such order-dual semirings are known, for instance:%
\begin{itemize}
\item The \textbf{completed max-times and min-times semifields.}
\begin{align}
\label{eq:pair:mmt}
\cmaxtimes &= \langle [0,\infty], \max, \timesl, \linv{\cdot}, 0, 1, \infty\rangle
&
\cmintimes &= \langle [0,\infty], \min, \timesu, \linv{\cdot}, \infty, 1, 0\rangle
\end{align}

\item The \textbf{completed max-plus (schedule algebra, polar algebra) and min-plus semifields (tropical algebra).}
\begin{align}
\label{eq:pair:mmp}
\cmaxplus &= \langle [-\infty,\infty], \max, \plusl, -{\cdot}, -\infty, 0, \infty\rangle
&
\cminplus &= \langle [-\infty,\infty], \min, \plusu, -{\cdot}, \infty, 0, -\infty\rangle
\end{align}
\end{itemize}
Note that their additions are all idempotent: a semiring with idempotent addition is simply called an \emph{idempotent semiring} and it is always positive. 
These find usage in path-finding algorithms, and some more examples can be found in~\cite{gon:min:08}.  
The mechanism whereby they are exposed as pairs of dually ordered semifields is explained in Section~\ref{sec:construction}. 
\qed
\end{Example}

%
\iftoggle{extendedVersion}{
\subsubsection{Semimodules over positive semifields}

Let $\mathcal D = \langle D, +, \times, \epsilon_D, e_D\rangle$ be a commutative semiring.  
A \emph{$\mathcal D$-semimodule} $\mathcal X = \langle X, \oplus, \odot, \epsilon_X\rangle$ is a commutative monoid $\langle X, \oplus, \epsilon_X \rangle$ endowed with a scalar action $(\lambda, x) \mapsto \lambda \odot x$\, satisfying the following conditions for all $\lambda, \mu \in D$, $x, x' \in X$:
\begin{align}
\label{def:smm}
(\lambda \times \mu)\odot x &= \lambda \odot (\mu \odot x)
&
\lambda \odot (x \oplus x') &= \lambda \odot x \oplus \lambda \odot x' \\
(\lambda + \mu)\odot x &= \lambda \odot x \oplus \mu \odot x \notag
&
\lambda \odot \epsilon_X &= \epsilon_X = \epsilon_D \otimes  x \notag
\\
e_D \odot x &= x \notag
\end{align}
Matrices form a $\mathcal D$-semimodule $D^{g\times m}$ for given $g$, $m$\,. 
In this paper, we only use finite-dimensional semimodules where we can identify semimodules with column vectors, e.g.  $\mathcal X \equiv \mathcal D^g$\,. 
If $\mathcal D$ is commutative, 
naturally-ordered 
or complete, then $\mathcal X$ is also commutative, 
naturally-ordered 
or complete~\cite{gol:03}.
If $\mathcal K$ is a semifield, we may also define an inverse for the semimodule by the coordinate-wise inversion, 
$(\linv x)_i = \linv{(x_i)}$.

%
%


Similarly, the may define a \emph{matrix conjugate} $\left (\lconj A\right)_{ij} = \linv A_{ji}$\,.  
For complete idempotent semifields, the following matrix algebra equations are proven in~\citep[Ch.8]{cun:79}:
\begin{proposition}
\label{prop:mat:eq}
\label{lem:mod:alt:prods}
Let $\mathcal K$ be an idempotent semifield, and $A \in \mathcal K^{m\times{n}}$. Then:
\begin{enumerate}
\item \label{prop:mat:eq:1}
$A \otimesl (\lconj A \otimesu A ) = A \otimesu (\lconj A \otimesl A) = %
(A \otimesu \lconj A ) \otimesl A = (A \otimesl \lconj A ) \otimesu A = A$ and 
$\lconj A \otimesl ( A \otimesu \lconj A ) = \lconj A \otimesu ( A \otimesl \lconj A) = %
(\lconj A \otimesu  A ) \otimesl \lconj A = (\lconj A \otimesl  A ) \otimesu \lconj A = \lconj A$\enspace.
\item \label{lem:mod:alt:prods:2} Alternating $A-\lconj A$ products of 4 matrices can be shortened as in:
\begin{align*}
\lconj A \otimesu (A \otimesl (\lconj A \otimesu A)) = \lconj A \otimesu A
= (\lconj A \otimesu A ) \otimesl (\lconj A \otimesu A)
\end{align*}
\item \label{lem:mod:alt:prods:3} Alternating $A-\lconj A$ products of 3 matrices and another terminal, arbitrary matrix can be shortened as in:
\begin{align*}
\lconj A \otimesu (A \otimesl (\lconj A \otimesu M)) = \lconj A \otimesu M
= (\lconj A \otimesu A ) \otimesl (\lconj A \otimesu M)
\end{align*}
\item \label{lem:mod:alt:prods:4} The following inequalities apply:
\begin{align*}
\lconj A \otimesu (A \otimesl M ) &\geq M
&
\lconj A \otimesl (A \otimesu M ) &\leq M
\end{align*}
\end{enumerate}
\end{proposition}

}{
}

\subsubsection{A construction for positive semifields}
\label{sec:construction}
There is a non-countable number of semifields obtainable from $\nnR$\,. Their discovery is probably due to Maslov and collaborators~\cite[\S1.1.1]{mas:vol:88}, but we present here the generalized procedure introduced by Pap and others~\cite{pap:93,pap:ral:98,mes:pap:99} that includes Maslov's results as a particular case. 
\begin{cons}[Pap's dioids and semifields]
\label{cons:pap:basic}
Let $\nnR$ 
be the semiring of non-negative reals, and 
consider a strictly monotone \emph{generator function} $g$ on an interval $[a,b] \subseteq [-\infty, \infty]$ with endpoints in $[0,\infty]$. 
Since $g$ is strictly monotone it admits an inverse $g^{-1}$, so set
\begin{enumerate}
\item the pseudo-addition, $u \oplus v = g^{-1}(g(u) \plusl (g(v))$
\item the pseudo-multiplication, $u \otimes v =  g^{-1}(g(u) \timesl (g(v))$
\item neutral element, $e = g^{-1}(1)$
\item inverse, $\lconj{x} = g^{-1}(\frac{1}{g(x)})$,
\end{enumerate}
Then,
\begin{enumerate}
\item if $g$ is strictly monotone and increasing increasing such that $g(a) = 0$ and $g(b) = \infty$, then  
a complete positive semifield whose order is aligned with that of $\nnR$ is:
$$\cK_g = \langle [a, b], \oplusl, \otimesl, \lconj{\cdot}, \bot=a, e, \top=b \rangle\,.$$

\item order-dually, if $g$ is strictly monotone and decreasing 
such that $g(a) = \infty$ and $g(b) = 0$, then  a complete positive semifield whose order is aligned with that of $\linv{\nnR}$ is
$$\lconj{\cK_g} = \langle [a, b], \oplusu, \otimesu, \lconj{\cdot}, \lconj\bot=b, e, \lconj\top=a \rangle\,.$$
\end{enumerate}
\end{cons}
\begin{proof}
See~\cite{pap:ral:98,pap:93} for the basic dioid $\cK_g$, and~\citep[p.~49]{gon:min:08} for the inverse operation and the fact that it is a semifield, hence a positive semifield. The description of the operations of $\lconj{\cK_g}$ is provided by Theorem~\ref{theo:order:duals}.
\end{proof}
\noindent
Note how in the $g$-calculus operations are always named with a ``pseudo-'' prefix, but it does not agree with Semiring Theory practice, hence we drop it. 
Also, the effect of the type of motonicity of $g$ is to impose the polarity of the extended operations. 
Remember that the inversion $\lconj{\cdot}$ is a dual isomorphism of semifields, so that $\lconj{\left( \lconj{\cK_g}\right)} = \cK_g$ and $\lconj{\left( \cK_g\right)} = \lconj{\cK_g}$\,. 
The different notation for the underlying inverse $\linv{\cdot}$ and the inverse in Pap's construction $\lconj{\cdot}$ is introduced so that it can later be instantiated in a number of constructed inverses, as follows. 

Our use of Construction~\ref{cons:pap:basic} is to generate different kind of semifields by providing different generator functions:
\begin{cons}[Multiplicative-product real semifields~\cite{mes:pap:99}]
\label{cons:plustimes}
Consider a free parameter $r \in [-\infty, 0)\bigcup (0, \infty]$ and the function $g(x) = x^r$ in $[a,b]=[0,\infty]$ in Construction~\ref{cons:pap:basic}.
For the operations we obtain:
\begin{align}
u \oplus_r v &= \left( u^r \plusl v^r \right)^\frac{1}{r}
&
u \otimes_r v &= \left (u^r \timesl v^r\right)^\frac{1}{r} = u \timesl v
&
\lconj u = \left (\frac{1}{u^r} \right)^\frac{1}{r} = \linv{u}
\end{align}
where the basic operations are to be interpreted in $\nnR$\,. 
Now, 
\begin{itemize}
\item if $r \in (0,\infty]$ then $g(x)=x^r$ is strictly monotone increasing whence $\bot_r = 0$, $e_r = 1$, and $\top_r = \infty$\,, and the complete positive semifield generated, order-aligned with $\nnR$, is:
\begin{align}
\label{eq:smf:times:pos}
\nnRr = \langle [0,\infty], \oplusl_r, \timesl, \linv{\cdot}, \bot_r= 0, e, \top_r = \infty\rangle
\end{align}

\item if $r \in [-\infty, 0)$ then $g(x)=x^r$ is strictly monotone decreasing whence $\bot_r = \infty$, $e_r = 1$, and $\top_r = 0$\,, and the complete positive semifield generated, order-aligned with $\linv{(\nnR)}$, or dually aligned with $\nnR$, is:
\begin{align}
\label{eq:smf:times:neg}
\nnRr[-r] & 
	= \linv{\nnRr} 
	= \left( \linv{\nnR}\right)_r
	= \langle [0,\infty], \oplusu_r, \timesu, \linv{\cdot}, \linv\bot_r= \infty, e, \linv\top_r = 0\rangle
\end{align}
\end{itemize}
\end{cons}
\begin{proof}
By instantiation of the basic case. See the details in~\cite{val:pel:16e}.
\end{proof}
Note that $\nnRr[1] \equiv \nnR$ and $\nnRr[-1] \equiv \linv{\nnR}$. 
This suggests the following Corollary:
\begin{Corollary}
$\nnRr$ and $\nnRr[-r]$ are inverse, completed positive semifields. 
\end{Corollary}

In particular, consider the cases:
\begin{Corollary}
\label{prop:plustimes:cases}
In the previous Construction~\ref{cons:plustimes}, 
\begin{align}
\label{prop:plustimes:cases:one}
\nnRr[1] &= \nnR 
&
\nnRr[-1]  &= \linv{\nnR}\\
\label{prop:plustimes:cases:infty}
\lim_{r\to\infty} \nnRr &= \cmaxtimes
&
\lim_{r\to-\infty}  \linv{\nnRr} &=\cmintimes 
\end{align}
\end{Corollary}
\begin{proof}
The proof of~\eqref{prop:plustimes:cases:one} by inspection. For~\eqref{prop:plustimes:cases:infty} see~\cite{mes:pap:99}.
\end{proof}

All these semifields have the same product, and the same ``extreme'' points, $\{0,1,\infty \}$\,. Their only difference lies in the addition. Sometimes, when only the product is important in an application, the addition remains in the background and we are not really sure in which algebra we are working on. 
Note also that instead of using the abstract notation for the inversion $\lconj{\cdot} $, since $\nnR$ is the paragon originating all other behaviour, we have decided to use the original notation for the inversion in the (incomplete) semifield.

The case where $r = 0$ deserves to be commented. To start with, note that for $q > 0$, $\nnRr[q]$ is aligned with $\nnR$ while  $\nnRr[-q]$ is aligned with $\dual{(\nnR)}$, therefore $\lim_{r\rightarrow 0^+}\nnRr$ and  $\lim_{r\rightarrow 0^-}\nnRr$ are endowed with opposite orderings.
Therefore the following lemma is not a surprise. 
\begin{Lemma}
$\nnRr[0]$ is not a semifield. 
\end{Lemma}
\begin{proof}
Recall, that on finite operands $u \oplus_r v = \left( u^r + v^r\right)^{1/r}$. Then we may generalize, due to the associativity of the operation to a vector $\vec v = \{v_i\}_i$ to  $\bigoplus_{r,i} v_i = \left( \sum_i v_i^r\right)^{1/r}$. 
Next, consider $\vec w, \vec x \in (0,\infty)^n$, and $q \in (0, \infty)$. From the properties of the means (see also~\cite[C.~2]{har:lit:pol:52}) , it is easy to prove that
$$\sqrt[-q]{\sum_{i=1}^n w_i x_i^{-q}} \leq \prod_{i=1}^{n} x_i^{w_i} \leq \sqrt[q]{\sum_{i=1}^n w_i x_i^{q}}\,.$$
By considering $w_i = 1$ and introducing the limits as $q \rightarrow 0$ we have
\begin{align}
\bigoplus_{0,i} v_i = \prod_{i} v_i
\end{align}
whence \emph{the additive and the multiplicative structures of $\nnRr[0]$ are the same}, so the structure is not even a semiring. 
Furthermore, when $u, v \in [0,\infty]$, then $\lim_{r \rightarrow 0^-} u \oplus_r v = u \timesu v$ whereas $\lim_{r \rightarrow 0^+} u \oplus_r v = u \timesl v$. 
Hence, in particular when, say $u = 0$ and $v = \infty$ we have $\infty = u \otimesu v \not < u \otimesl v = 0$ and the product is  clearly not defined. 
\end{proof}


%
%
%
%

\section{Results}
\label{sec:results}
We are now ready to start presenting a chain of results that leads to our conjecture.

\subsection{Entropic semifields}
\label{sec:entropy:smf}

\subsubsection{The basic entropic semifield}
\label{sec:basic:entsmf}
The effect of Hartley's information function is to induce from the set of positive numbers (restricted to the $[0,1]$ interval) a semifield of the extended reals $[0,\infty]$\,. 
To see this, we actually consider it acting on the whole of the non-negative reals $\nnR$ of \eqref{eq:pair:nnR} onto the algebra of entropies denoted by $\mathbb H$. 
%
\begin{Theorem}[Hartley's semifields]
\label{theo:hartley:smf}
The algebra 
$\langle [-\infty, \infty], \oplus, \otimes, \lconj{\cdot},  \infty, 0\rangle$
with
\begin{align}
\label{eq:h:ops}
h_1 \oplus h_2 &= h_1 + h_ 2 - \ln (e^{h_1} + e^{h_2})
&
h_1 \otimes h_2 &= h_1 + h_2 
&
\lconj{h} &=  -h 
\end{align}
obtained from that of positive numbers by Hartley's information function is a positive semifield that can be completed in two different ways to two mutually dual semifields: 
\begin{align}
\hR &= \langle [-\infty, \infty], \oplusl, \otimesl, -{\cdot}, \bot=-\infty, e=0 , \top=\infty\rangle
&
-{\hR} &= \langle [-\infty, \infty], \oplusu, \otimesu, -{\cdot},  -\bot=\infty, e=0 , -\top=-\infty\rangle
\end{align}
whose elements can be considered as entropic values and operated accordingly. 
\end{Theorem}
\begin{proof}
Define the extension to Hartley's information function to non-negative numbers $\hinf{\cdot}: [0,\infty] \rightarrow [-\infty, \infty]$ as $\hinf{p}= - \ln p$. This is  one-to-one from $[0,\infty]$ and total onto $[-\infty, \infty]$, with inverse $\hinfinv{h} = e^{-h}$ for $h \in [-\infty, \infty]$.  

Since $\hinfinv{h} = e^{-h}$ is monotone, it is a generator (function) for Construction~\ref{cons:pap:basic}, with  the following addition, multiplication and inversion:
\begin{align*}
h_1 \oplus h_2 &= \hinf{ \hinfinv{h_1} + \hinfinv{h_2}} 
 = - \ln \left( e^{- h_1} +  e^{-h_2}\right) 
 = \ln \frac{e^{h_1 + h_2}}{e^{h_1} + e^{h_2}} = h_1 + h_ 2 - \ln \left(e^{h_1} + e^{h_2}\right)
 \\
h_1 \otimes h_2 &= \hinf{\hinfinv{h_1} \times \hinfinv{h_2}}
 = - \ln \left( e^{- h_1}\cdot e^{-h_2}\right) = - \ln\left( e^{-(h_1 + h_2)}\right) = h_1 + h_2 
 \\
\lconj{h} &= \hinf{ \frac{1}{\hinfinv{h}}} = - \ln\frac{1}{e^{-h}} = - h\,.
\end{align*}
The ``interesting'' points $\{0,e,\infty\}$ are transformed as: 
\begin{align}
\hinf{0} &= \infty 
&
\hinf{1}  &= 0 
&
\hinf{\infty} &= -\infty 
\end{align}
The rest follows by Construction~\ref{cons:pap:basic} and Theorem~\ref{theo:order:duals}.
\end{proof}
Several considerations are worth stating. First, we have not restricted this dual-order isomorphism to the sub-semiring of probabilities on purpose. 
However, on $\hinf{[0,1]}  = [0,\infty] \subseteq [-\infty, \infty]$ our intuitions about amounts of informations hold; we still believe that the ``information'' in $\mathfrak I^\ast(0) = \infty$ is the highest, whereas the probability of $\left (\mathfrak I^\ast\right)^{-1}(\infty)=0$ is the smallest. 

Second, the notation for the inverse of Hartley's semifield $-\hR$ is a mnemonic to remind that this is a semifield in which the inversion is actually an \emph{additive} inverse, and consequently the product is an \emph{addition}.

Third, the order of $\nnR$ is aligned with that of  $\hR$, actually a sub-order of it in the abstract algebra sense: same properties on a subset of the original carrier set. But the order of $-{\hR}$ is aligned with the \emph{dual} order,  that of $\linv{\nnR}$. 
This has to be taken into consideration in the application of Theorem~\ref{theo:order:duals} to the proof of Theorem~\ref{theo:hartley:smf}. 
Corollary~\ref{coro:hartleys} makes this difference clear. 
\begin{Corollary}
\label{coro:hartleys}
Hartley's information function extended to the non-negative reals $\hinf{m} = - \log_b m = h \leftrightarrow \hinfinv{h} = b^{-h}= m$ is a dual-order isomorphism of completed positive semifields between $\nnR$ and $-{\hR}$\,.  
\end{Corollary}
\begin{proof}
In the previous theorem, note that $\hinf{\cdot}$ is monotonically \emph{decreasing}, entailing that the construction inverts orders in semifields, that is $\hinf{\nnR} = -{\hR}$ and $\hinf{\linv{\nnR}} = \hR$. 
\end{proof}

\begin{figure}
\centering
\centering
\begin{tikzpicture}[
	scale=1.0,
	adomain/.style={circle, draw,fill=white,radius=0.1cm},
	nodomain/.style={circle, draw,dashed,radius=0.2cm},
	label distance=2mm,
	]
	\tikzmath{
		integer \lowaxis, \highaxis, \mInf, \pInf, \midaxis;
		\lowaxis = 0; 	
		\highaxis = 2; 
		\midaxis = (\lowaxis + \highaxis)/2;
		\mInf = -6; 	
		\pInf = 6;		  
		\rcoord = 3;    
	}
	\path[use as bounding box] (\mInf-0.5,\lowaxis-1.5) rectangle (\pInf+0.5,\highaxis + 1.5); 

	\draw[->] (\mInf,\highaxis) -- (\pInf,\highaxis);
	\draw[->] (\mInf,\lowaxis) -- (\pInf,\lowaxis);
	\node[nodomain] (R0) at (0,\lowaxis) {} node [above = 0.15cm of R0] {$r=0$};
	\node[nodomain] (H0) at (0,\highaxis) {};
	
	\node[adomain] 	(Rbot)	at (\mInf-0.2,\lowaxis) {} node [below of=Rbot] {$\nnRr[-\infty]$};
	\node[adomain] 	(Rm1)	at (-1,\lowaxis) {} node [below of=Rm1] {$\nnRr[-1]$};
	\node[adomain] 	(R1)	at (1,\lowaxis) {} node [below of=R1] {$\nnRr[1]\equiv \nnR$};
	\node[adomain] 	(Rr)	at (\rcoord,\lowaxis) {} node [below of=Rr] {$\nnRr[r]$};
	\node[adomain] 	(Rtop)	at (\pInf+0.2,\lowaxis) {} node [below of=Rtop] {$\nnRr[\infty]$};
	
	\node[adomain] 	(Htop)	at (\mInf-0.2,\highaxis) {} node [above of=Htop] {$\hRr[-\infty] = \cmaxplus$};
	\node[adomain] 	(Hm1)	at (-1,\highaxis) {} node [above of=Hm1] {$\hRr[-1]$};
	\node[adomain] 	(H1)	at (1,\highaxis) {} node [above of=H1] {$\hRr[1]\equiv -\hR$};
	\node[adomain] 	(Hr)	at (\rcoord,\highaxis) {} node [above of=Hr] {$\hRr[r]$};
	\node[adomain] 	(Hbot)	at (\pInf+0.2,\highaxis) {} node [above of=Hbot] {$\hRr[\infty] = \cminplus$};
	
	\draw[->]  (Hr.south east) to [bend left]  node [right] {$\varphi'_r$} (Rr.north east) ;
	\draw[->]  (Rr.north west) to [bend left]  node [left] {$(\varphi'_r)^{-1}$} (Hr.south west) ;
	\draw[->]  (Hbot.south east) to [bend left]  node [right] {$\varphi'_{\infty}$} (Rtop.north east) ;
	\draw[->]  (Rtop.north west) to [bend left]  node [left] {$(\varphi'_{\infty})^{-1}$} (Hbot.south west) ;
	\draw[->]  (Htop.south east) to [bend left]  node [right] {$\varphi'_{-\infty}$} (Rbot.north east) ;
	\draw[->]  (Rbot.north west) to [bend left]  node [left] {$(\varphi'_{-\infty})^{-1}$} (Htop.south west) ;

\end{tikzpicture}
\caption{\textbf{Domain diagram to interpret the R\'enyi transformation in the context of the expectations of probabilities and informations.} }
\label{fig:domain:diagram}
\end{figure}
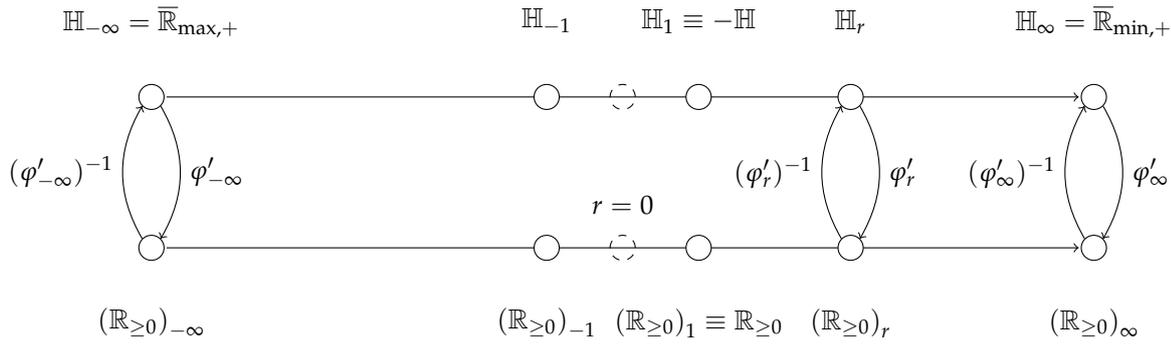

\subsubsection{Constructed entropic semifields}
\label{sec:built:entsmf}

The next important fact is that R\'enyi's modified averaging function $\varphi(p) = \frac{-1}{r}\log_b p$ and its inverse are also dual isomorphisms of positive semifields. 
To prove this we make said functions appear in the construction of the semifields as generators. 
\begin{Theorem}[Additive-product real semifields or Entropy semifields]
\label{theo:entropy:smf}
Let $r \in [-\infty, \infty] \setminus \{0\}$ and $b \in (1,\infty)$. 
Then the algebra $\langle [-\infty, \infty], \oplus_r, \otimes_r, \linv{\cdot}, \bot= \infty, e= 0 \rangle$ 
whose basic operations are:
\begin{align}
u \oplus_r v &= 
u + v - \log_b \left( b^{ru} + b^{rv}  \right)^{\frac{1}{r}}
&
u \otimes_r v & 
= u +  v
&
\lconj u 
		= - {u}
\end{align}
can be completed to two dually-ordered positive semifields
\begin{align}
\hRr &= \langle [-\infty, \infty], \oplusl_r, \otimesl_r, -{\cdot}, \bot=-\infty, e=0 , \top=\infty\rangle
\\
-{\hRr} &= \langle [-\infty, \infty], \oplusu_r, \otimesu_r, -{\cdot},  -\bot=\infty, e=0 , -\top=-\infty\rangle
\end{align}
whose elements can be considered as emphasized, entropic values and operated accordingly. 
\end{Theorem}
\begin{proof}
We build these semifields with a composition of results.  The first one is the well known result from the theory of functional means we choose to cast into the framework of Pap's $g$-calculus: the power mean of order $r$ is the pseudo arithmetic-mean with generator $\varphi_r(x) = x^r$ and inverse $\varphi_r^{-1}(y) = y^{1/r}$. This was used in Construction~\ref{cons:plustimes} to build the semifields of \eqref{eq:smf:times:pos} and \eqref{eq:smf:times:neg}. The second result is Corollary~\ref{coro:hartleys}, where $\hinf{\cdot}$ is proven a dual order isomorphism of semirings. 

We next use the composition of functions $\varphi_r' = \mathfrak I_\ast  \circ \varphi_r$ and its inverse $\linv{(\varphi'_r)} = \linv{\varphi_r} \circ \linv{\mathfrak I_\ast}$. The latter exists, since it is a composition of isomorphisms and it is a dual order isomorphism, since $\hinf{\cdot}$ is order-inverting while $\varphi_r$ is not. 
That composition is precisely R\'enyi's  function $\varphi_r'(h)  =  b^{-rh}$ with inverse $\linv{(\varphi'_r)}(p) = \frac{-1}{r}\log_b p$\,, whence: 
\begin{align*}
u \oplus_r v &=\linv{(\varphi'_r)}(\varphi'_r(u) + \varphi'_r(v)) 
  = \frac{-1}{r}\log_b\left( b^{-ru} + b^{-rv}\right)
  = \frac{-1}{r}\log_b\left( \frac{b^{ru} + b^{rv}}{b^{r(u + v)}}\right)
  = u + v - \log_b \left( b^{ru} + b^{rv}  \right)^{\frac{1}{r}}
\\
u \otimes_r v &= \linv{(\varphi'_r)}(\varphi'_r(u) \times \varphi'_r(v)) 
  = \frac{-1}{r}\log_b\left( b^{-ru} \times  b^{-rv}\right)
  = \frac{-1}{r}\log_b\left( b^{-r(u + v)}\right) = u + v
\\
\lconj u &= \linv{(\varphi'_r)}\left(\frac{1}{\varphi'_r(u)}\right) = \frac{-1}{r}\log_b\left( \frac{1}{b^{-ru}}\right) = - u
\end{align*}
This composition is strictly increasing when $r \in [-\infty, 0)$ and strictly decreasing when $r \in (0, \infty]$, hence, when applying Construction~\ref{cons:pap:basic} with it: 
\begin{itemize}
\item for $r \in (0, \infty]$ we obtain  
$-{\hRr} = \hinf{\nnRr} = \langle [-\infty, \infty], \oplusu_r, \otimesu_r, -{\cdot},  - \bot=\infty, e=0 , -\top=-\infty\rangle$\,, and  
\item for $r \in [-\infty, 0)$  we obtain 
	$\hRr = \hinf{\linv{\nnRr}} = \langle [-\infty, \infty], \oplusl_r, \otimesl_r, -{\cdot}, \bot=-\infty, e=0 , \top=\infty\rangle$\,,
\end{itemize}
with the extended operations:
\begin{align*}
-\bot &= -(-\infty)  = \infty = \top
&
-\top &= - \infty = \bot
\\
u \oplusl_r v & = 
\begin{cases}
\infty & \text{if } u = \infty  \text{ or } v = \infty \\
u \oplus_r v & \text{otherwise}
\end{cases}
&
u \otimesl_r v & =
\begin{cases}
-\infty & \text{if } u = -\infty \text{ or } v = -\infty\\
u + v & \text{otherwise}
\end{cases}
\\
u \oplusu_r v & =
\begin{cases}
- \infty & \text{if } u = - \infty\text{ or } v = - \infty\\
u \oplus_r  v & \text{otherwise}
\end{cases}
&
u \otimesu_r v & =
\begin{cases}
\infty & \text{if } u = \infty \text{ or } v = \infty\\
u + v & \text{otherwise}
\end{cases}
\end{align*}
\end{proof}
Note that in these semirings we have:
$-\infty \otimesl_r \infty = -\infty$ and $-\infty \otimesu_r \infty = \infty$ for $r \in [-\infty, \infty]\setminus\{0\}$\,. 
Moreau's original proposals~\cite{mor:70}  are found for $r=1$ $-\infty \timesl \infty = -\infty$ and $-\infty \timesu \infty = \infty$. 
Further details for compositions of generating functions and other averaging constructions can be found in~\cite{gra:mar:mes:pap:11b}.
%

The following corollary is the important result we announced at the beginning of this section. 
\begin{Corollary}
The R\'enyi entropies in the R\'enyi spectrum $\srentropy[r]{P_X}$ take values in the semifields  $\hRr$.
\end{Corollary}
\begin{proof}
We notice that the generating function to activate Construction~\ref{cons:pap:basic} in Theorem~\ref{theo:entropy:smf} 
 is none other than the function to calculate the R\'enyi non-linear average $\varphi'$ of~\eqref{eq:adapting}. 
 Hence the values resulting from R\'enyi's entropies belong in that semifield of Theorem~\ref{theo:entropy:smf} with the respective $r$ parameter.
\end{proof}

Similar to Proposition~\ref{prop:plustimes:cases} we have the following:
\begin{Proposition}
\label{prop:someplus:cases}
In the semifield construction  of Theorem~\ref{theo:entropy:smf}
\begin{align}
\label{prop:someplus:cases:one}
\hRr[-1] &= \hR
&
\hRr[1]  &= -{\hR}\\
\label{prop:someplus:cases:infty}
\lim_{r\to -\infty} \hRr &= \cmaxplus
&
\lim_{r\to \infty}  \hRr &=\cminplus
\end{align}
\end{Proposition}
\begin{proof}
The proof of~\eqref{prop:someplus:cases:one} by inspection. For~\eqref{prop:someplus:cases:infty} see~\cite{gon:min:08,mes:pap:99}.
\end{proof}
The apparent incongruity of $\hRr[1] = -\hR$ stems from the different origins of each notation: $\hRr[1]$ comes from the theory of semirings and is due to the fact that we have forced it to be aligned with $-\hR$ which is ultimately motivated by Shannon's choice of sign for the entropy function~\cite{sha:48a}. Note that in Thermodynamics, this choice of sign is the inverse, that is,  \emph{negentropy} is the privileged concept.


%
\subsection{Application: rewriting entropies in semifields}
\label{sec:rewriting:entropy}
We would like to clarify the meaning of these many semifields in relation to the entropy. Due to Definition~\ref{def:sRenyie} 
one way to do so would be to rewrite the means in \eqref{eq:holder:wmean} 
in terms of the adequate semifields. After that, all of the entropic concepts would be written in terms of the semifield-expressed generalized weighted means. 
For that purpose, we will follow the painstakingly developed description of the means in~\cite[Chap.~II]{har:lit:pol:52}, by writing $\shmean{\vec w}{\vec x}$ for the mean rewritten in the complete semifield notation. 

We propose the following definition. 
%
%
\begin{Definition}
\label{lemma:means:in:smf}
Let $\vec x, \vec w \in [0,\infty]^n$ and $r\in [-\infty,\infty]$\,. Then in the dual completed semifields issuing from $\nnR$, 
\begin{align}
\label{eq:means:in:smf}
\shmean{\vec w}{\vec x} &= 
\begin{cases} 
\left ( \bigoplusu_i  \left( w_i \overu \bigoplusu_k w_k \right) \otimesu x_i^r \right)^\frac{1}{r} & r < 0\\
e^{\shmean[1]{\vec w}{\ln{\vec x}}} & r = 0\\
\left ( \bigoplusl_i  \left( w_i \overl \bigoplusl_k w_k \right) \otimesl x_i^r \right)^\frac{1}{r} & r > 0
\end{cases}
\end{align}
where the case for $r=0$ is only valid when there are no $i \neq j$ with $x_i^{w_i} = 0$ and $x_j^{w_j} = \infty$\,.
\end{Definition}
%
%
\begin{proof}[Justification] 
To justify this definition, 
we will carry out a case-based analysis based in the following set definitions: Let $n$ be the dimension of $\vec x$ and $\vec w$ and $\bar n = \{1, \ldots, n \}$ the full set of indices on all the components, and let: 
\begin{align}
\xbot  &= \{ i \in \bar n \mid x_i = 0 \}
&
\xtop &= \{ i \in \bar n \mid  x_i = \infty \} 
&
\xfin &= \{ i \in \bar n \mid x_i \in (0, \infty)\} 
\end{align}
be, respectively, the set of indices of zero, finite, and infinite components, and notice that $\bar n = \xbot \cup \xfin \cup \xtop$. 

Note that the corner cases when $\vec x = 0$ 
are catered to by the reflexivity of the means~\cite{har:lit:pol:52}. 
In particular, when $r > 0$, if $\vec x = 0$, then $\xbot = \bar n$ and we require that $\shmean{\vec w}{\vec x} = 0$. 
This entails that the multiplication in~\eqref{eq:means:in:smf} for $r >0 $ must be the lower multiplication so that
$$
\shmean{\vec w}{\vec x} = \left ( \bigoplusl_i  \srdist[0]{\vec w} 
\otimesl 0 \right)^\frac{1}{r} = 0
$$

In~\cite[Chap.~II]{har:lit:pol:52} the existence of $w_i = 0$ is directly disallowed, and, we take the same is the case for $w_i = \infty$. However, the authors later take into consideration the second possibility, so we have decided to include both from the start. 
For this purpose, consider the case where $\xbot[\vec w] = \bar n$. 
%
Let us first define $W = \bigoplusl_i w_i$, which is the only possible definition consistent with our intuitions of an overall weight $W$ for $ r > 0$, and notice that $\xbot[\vec w] = \bar n \iff W = 0 \iff \lconj W = \infty$. 
This means we could, in principle, 
define the normalized weight distribution $\srdist[0]{\vec w}$ in two possible  ways:
\begin{enumerate}
\item either $\srdist[0]{\vec w} = \{ w_i \overl W\} = \{ w_i \otimesu\lconj W  \}$; 
\item or $\srdist[0]{\vec w} = \{ w_i \overu W\} = \{ w_i \otimesl \lconj W\}$. 
\end{enumerate}

In the first case, if all the weights are null, that is $\xbot[\vec W] = \bar n$ then $ \forall i, w_i \otimesu \lconj W = 0 \otimesu \infty = \infty$. For zero coordinates of $\vec x$ this poses no problem, but for $j \in \overline{\xbot} \neq \varnothing$ 
 we find that $(w_j \otimesu \lconj W) \otimesl x_j^r = \infty \otimesl x_i^r = \infty$ whence the whole summation is infinite. This does not seem reasonable. 
On the other hand, in the second case, we have that $(w_i \otimesl \lconj W) \otimesl x_i^r = (0 \otimesl \infty) \otimesl x_i^r =0 \otimesl x_i^r = 0$, as expected. 

However, this has too much ``annihilating power'' for if 
$\xtop[\vec w] \neq \varnothing \iff W = \infty \iff \lconj W = 0$, considering an  $ i \in \xtop[\vec w]$ such that $w_i = \infty$, then  \emph{every  factor  is erased} since for $j \in \overline{\xbot[\vec w]}$ is
$(w_j \otimesl \lconj W) \otimesl x_j^r = (w_j \otimesl 0) \otimesl x_j^r = 0$, 
and the alternative is even less intuitive than the preceding one. 
Therefore we are forced to choose alternative 1 $\srdist[0]{\vec w} = \{ w_i \overl W\}$ with the caveat that when all of the weights are null (a strange situation indeed) this does not make sense. 

Two points are worth stating:
\begin{itemize}
\item The case where $r <0$ 
is reasoned out by duality, with $\xtop$ being dual to $\xbot$, $r <0$ being the dual to $r >0$,  $\infty$ dual to $0$, and the \emph{upper} multiplication and addition  duals to the \emph{lower} multiplication and addition. In the following we just use this "duality" argument to solve the case for $r < 0$. 

\item The rest of the cases to analyze essentially have $\xbot \neq \bar n$ and $\xbot[\vec w] \neq \bar n$ whence their complements  are non-null $\overline{\xbot} \neq \varnothing, \overline{\xbot[\vec w]} \neq \varnothing$. This means that the actual summation in~\eqref{eq:means:in:smf} is extended to $\overline{\xbot} \cap \overline{\xbot[\vec w]}$, with $\overline{\xbot} = \xfin \cup \xtop$ and $\overline{\xbot[\vec w]} = \xfin[\vec w] \cup \xtop[\vec w]$. 

\end{itemize}

Due to Theorem~\ref{theo:order:duals} and Construction~\ref{cons:pap:basic} we know that  if $\xfin = \overline{\xbot} $ and $\xfin[\vec w] = \overline{\xbot[\vec w]}$---that is all non-zero coordinates are finite%
---then the expressions reduce to those of the classical definition in~\eqref{eq:holder:wmean}  so that $\shmean{\vec w}{\vec x} =  \hmean{\vec w}{\vec x}$ for $r \in [-\infty,\infty]$, that is including $r\leq 0$. 

The only cases left to analyze for $ r > 0$ are those where $\infty$ appears in either the weights $\vec w$ or the quantities $\vec x$ on subindices $i \in \overline{\xbot} \cap \overline{\xbot[\vec w]}$. 
Notice that if $i \in \xtop$ then $(w_i \otimesu \lconj W) \otimesl \infty = \infty$ since $w_i \otimesu \lconj W$ cannot be zero, therefore $\shmean{\vec w}{\vec x} = \infty$. 
Likewise, if $i \in  \xtop[\vec w]$ then $W = \infty \iff \lconj W = 0$ and $w_i \otimesu \lconj W = \infty \otimesu 0 = \infty$ whence the factor $(w_i \otimesu \lconj W) \otimesl x_i^r = \infty \otimesl x_i^r  = \infty$ and $\shmean{\vec w}{\vec x} = \infty$. 
We have collected all cases for $r > 0$, in Table~\ref{fig:shmean:cases}, along with those for $ r < 0$ obtained by duality. 
\begin{table}
\begin{subfigure}[b]{0.5\textwidth}
\centering
\renewcommand{\arraystretch}{2}
\begin{tabular}{|l|c|c|c|c|} \cline{3-5}
\multicolumn{2}{c|}{\multirow{2}{*}{$\shmean[r < 0]{\vec w}{\vec x}$}} & \multirow{2}{*}{$\xtop = \bar n$ } & \multicolumn{2}{c|}{$\overline{\xtop} \neq \varnothing$} \\\cline{4-5}
\multicolumn{2}{c|}{} &  & $\xfin = \bar n$ & $\xbot \neq \varnothing$ \\
\hline
\multicolumn{2}{|c|}{$\xtop[\vec w] = \bar n$} & $\infty$ & $0$ & $0$\\
\hline
\multirow{2}{*}{\rotatebox{90}{$\overline{\xtop[\vec w]} \neq \varnothing$}}& $\xfin[\vec w]  = \bar n$ & $\infty$ & $\hmean{\vec w}{\vec x}$ & $0$\\
\cline{2-5}
& $\xbot[\vec w]  \neq \varnothing$  & $\infty$ & $0$ & $0$\\
\hline
\end{tabular}
\caption{For $r <0$}
\label{fig:shmean:cases:negr}
\end{subfigure}
\begin{subfigure}[b]{0.5\textwidth}
\centering
\renewcommand{\arraystretch}{2}
\begin{tabular}{|l|c|c|c|c|} \cline{3-5}
\multicolumn{2}{c|}{\multirow{2}{*}{$\shmean[r > 0]{\vec w}{\vec x}$}} & \multirow{2}{*}{$\xbot = \bar n$ } & \multicolumn{2}{c|}{$\overline{\xbot} \neq \varnothing$} \\\cline{4-5}
\multicolumn{2}{c|}{} &  & $\xfin = \bar n$ & $\xtop \neq \varnothing$ \\
\hline
\multicolumn{2}{|c|}{$\xbot[\vec w] = \bar n$} & $0$ & $\infty$ & $\infty$\\
\hline
\multirow{2}{*}{\rotatebox{90}{$\overline{\xbot[\vec w]} \neq \varnothing$}}& $\xfin[\vec w]  = \bar n$ & $0$ & $\hmean{\vec w}{\vec x}$ & $\infty$\\
\cline{2-5}
& $\xtop[\vec w]  \neq \varnothing$  & $0$ & $\infty$ & $\infty$\\
\hline
\end{tabular}
\caption{For $r > 0$}
\label{fig:shmean:cases:posr}
\end{subfigure}
\caption{A summary of cases for $\shmean{\vec w}{\vec x} $ a) for $r <0$ and b) for $r > 0$}
\label{fig:shmean:cases}
\end{table}

Only the case $r=0$ is left for analysis, and recall this is $\hmean[0]{\vec w}{\vec x} = \left (\Pi_i x_i^{w_i} \right)^{\frac{1}{\sum_k w_k}} $. 
In~\cite{har:lit:pol:52} this case is treated exceptionally throughout the treatise, and 
one of the often used alternatives expressions for it is $\hmean[0]{\vec w}{\vec x} = \exp\{\sum_i \frac{w_i}{W}\ln{x_i}\}$
leading immediately to $\hmean[0]{\vec w}{\vec x} = \exp\{\hmean[1]{\vec w}{\ln{\vec x}}\}$ 
where the logaritm has to be interpreted entry-wise. 

Recalling that $\log(\cdot)$ and $\exp(\cdot)$ are order isomorphisms of semifields, this suggests that we are trying to solve the hard problem for $r=0$ in the semifield $\hR$ by using $r=1$ thanks to the properties of the logarithm. 
However, the addition in $\hR$ is a \emph{multiplication} so the rules for calculating the weighted means there might be different, considering that  $\log x_i$ might be negative, or,  specially, that factors like $w_i \cdot \log x_i$ are actually $\hR$ \emph{exponentiations}, i.e. they come from factors $\log(x_i^{w_i})$. 

In fact, the clue to suggest the form for this mean comes from information theory and the requirement that $0 \cdot \log\frac{1}{0} = 0$, necessary to ignore impossible events with $p_i = 0$ with $- p_i \cdot \log p_i = 0$ in the entropy calculation. 
This can be modeled in our framework by demanding that this multiplication be $0 \otimesl \log({1} \overl {0}) = 0 \otimesl \infty = 0$ and to write the geometric mean as:  
\begin{align}
\label{def:geo:mean}
\shmean[0]{\vec w}{\vec x} &= \exp\{\shmean[1]{\vec w}{\ln{\vec x}}\} 
\end{align}

Note that since the logarithm changes the base semifield for $\shmean[1]{\vec w}{\log{\vec x}}$ the Table~\ref{fig:shmean:cases}.\subref{fig:shmean:cases:posr} has to be reinterpreted, along with the sets of subindices: if the components of a vector $\vec x$ belong to a semifield $\mathcal K$ with carrier set $K$, then the index sets have to be defined with respect to the semifield in it. Since the carrier set of $\hR$ is $\cR = [-\infty, \infty]$ then they read as:
\begin{align*}
\xbot  &= \{ i \in \bar n \mid x_i = \bot_{\cR} = - \infty \}
&
\xtop &= \{ i \in \bar n \mid  x_i = \top_{\cR} = \infty \} 
&
\xfin &= \{ i \in \bar n \mid x_i \in (\bot_{\cR }, \top_{\cR})\} 
\end{align*}
So a table similar to  Table~\ref{fig:shmean:cases}.\subref{fig:shmean:cases:posr}  for $\shmean[1]{\vec w}{\log{\vec x}}$ would have $-\infty$ instead of $0$. 
Nevertheless, the exponentiation would bring the mean back to $\nnR$, and entailing that \eqref{def:geo:mean} 
actually follows 
Table~\ref{fig:shmean:cases}.\subref{fig:shmean:cases:posr}. 
\end{proof}

%

%
%

This concludes our justification of the casting of the weighted means into semifield algebra. 
Note that a single $\vec x$ may have  $x_i = 0$ and $x_j = \infty$  for $i \neq j$, a case explicitly addressed in~\cite{har:lit:pol:52}: 
$\shmean{\vec p}{\vec x} = \infty \otimesl u(r)$ where $u(r)$ is the step function, undefined at $r=0$
$$u(r) = \begin{cases}
0 & r < 0 \\
1 & r > 0
\end{cases}
$$

With this formulation we are now capable of describing the R\'enyi entropies in semifield notation: 
\begin{Definition}
Let  $P_X(x_i) = p_i$ and $Q_X(y_i) = q_i$ be two distributions with compatible support. Then in the dual completed semifields issuing from $\nnR$, 
the expression of the R\'enyi cross-entropy, entropy and divergence are: 
\begin{align*}
\srentropy{P_X} &= - \log {\shmean{P_X}{P_X}} 
\\
\srcrossent{P_X}{Q_X}  &= - \log \shmean{P_X}{Q_X}
\\
\srdiv{P_X}{Q_X} & = \log \shmean{P_X}{\frac{P_X}{Q_X}}
\end{align*}
\end{Definition}
Notice that  when $\xfin[P_X] = \bar n$ and $\xfin[Q_X] = \bar n$ the definition of entropy based on the mean and in the semifield expressed are the same, whereas in the other cases there is an extension in the definition which is in agreement with several arbitrary choices made in the definition of the new means, e.g. $0 \otimesl\log 0 = 0$ to comply with the convention extant for entropies. 
This supports our claim that entropies can be written and operated in semifields. 

Since the equivalent probability function is a mean and the means can be expressed in a complete semifield, we have an expression of the former in the complete semifields of reals. 
Therefore, the expressions for the equivalent probability function and the information potential are:
\begin{align}
\srprob[r]{P_X} &= \shmean{P_X}{P_X} 
&
\srpot{P_X} &= \shmean{P_X}{P_X}^r
\end{align}
However, in the definition of the equivalente probability function, and the entropies, we have that $w_i = x_i$ which entails some of the cases in Figure~\ref{fig:shmean:cases} are not visited.

\subsection{Discussion: a conjecture on the abundance of semifields in Machine Learning and Computational Intelligence applications}
\label{sec:discussion}
We are now in a position to argue our conjecture about the abundance of semifields in knowledge domains that model intelligent behaviour. 
\begin{enumerate}
\item 
First, the shifting in definition of the R\'enyi entropy  by $r = \alpha -1$ in~\cite{val:pel:19a} leads to a a straightforward relation~\eqref{eq:sRenyE:entropies} between the power means of the probability distribution and the shifted R\'enyi entropy. 
For a given probability function or measure $P_X$ the evolution of  entropy with $r \in [-\infty, \infty]$ resembles an \emph{information spectrum} $\srentropy{P_X}$. 
In a procedure reminiscent of defining an inverse transform, we may consider an equivalent probability $\srprob{P_X}= b ^{-\srentropy{P_X}}$ ,which is the H\"older path of $P_X$, $\srprob{P_X} = \hmean[r]{P_X}{P_X}$\,.

\item 
The function used by R\'enyi to define the generalized entropy, when shifted, is the composition of two functions: Hartley's information function and the power function of order $r$, which are monotone and invertible in the extended non-negative reals $[0,\infty]$. 
They are also bijections:
\begin{itemize}
\item The power function is a bijection over of the extended non-negative reals, and 
\item  Hartley's is a bijection between the extended reals and the extended non-negative reals. 
\end{itemize} 

\item 
But in Construction~\ref{cons:pap:basic} both the power function and Hartley's prove to be isomorphisms of positive semifields, which are semirings whose multiplicative structure is that of a group, while the additive structure lacks additive inverses. 
Positive semifields are all naturally ordered and the power function respects this order within the non-negative reals, being an order isomorphism for generic power $r$. Importantly, positive semifields come in dually-ordered pairs and the expressions mixing operations from both members in the pair are reminiscent of boolean algebras. 

\begin{enumerate}
\item The power function $g(x) = x^r$ with $r\in [-\infty, \infty]\setminus\{0\}$ actually generates  a whole family of semifields $\nnRr$ related to emphasizing smaller (with small $r$) or bigger values (with big $r$) in the non-negative reals $\nnR$. Indeed, the traditional weighted means are explained by the Construction~\ref{cons:plustimes} as being power-deformed aritmetic means, also known as Kolmogorov-Nagumo means with the power function as generators. These, semirings come in dually-ordered pairs for orders $r$ $\nnRr$ and $-r$ $\nnRr[-r]$ whose orders are aligned or inverted with respect to that of $\nnR$. Indeed, $\nnR \cong \nnRr[1]$. 

\item However, Hartley's function is a dual-order isomorphism, entailing that the new order in the extended reals is the opposite of that on the non-negative reals. It actually mediates between the (extended) probability semifield $\nnR$ and the semifield of informations, notated as a homage to Hartley as $-\hR$. 

\end{enumerate}

\item 
Since the composition of the power mean and Hartley's information function produces 
the function that R\'enyi used for defining  his information measures, and this is a dual-order semifield isomorphism, being the composition of one dual-order isomorphism---Hartley's function---and an order isomorphism---the power function---we can see that entropies are actually operated in modified versions of Hartley's semifields $\hRr$ which come in pairs, as all completed positive semifields do. 

\item 
Many of the $\nnRr$ and $\hRr$ semifields appear in domains that model intelligent behaviour. Among a list of applications we list the following: 
\begin{itemize}
\item In AI, maximizing utilities and minimizing costs is used by many applications and algorithms, e.g. heuristic search, to mimic ``informed'' behaviour~\cite[Ch.~3]{rus:nor:10}, decision theory~\cite[Ch.~16]{rus:nor:10}, uncertainty and probability modelling~\cite[Ch.~13-15]{rus:nor:10}. In most applications $\nnRr[1]$, for multiplicatively-aggregated costs and utilities, and $\hRr[1]$, for additively aggregated ones are being used. 
Note that both a semifield and its order-dual are needed to express mixed utility-cost expressions, as in Electrical Network Analysis with resistances and conductances. 
%
Sometimes the idempotent versions of these spaces $\lim_{r \to \pm \infty}\nnRr$, e.g.$\cmintimes$ and  $\cmaxtimes$, and $\lim_{r \to \pm \infty}\hRr$, e.g. $\cminplus$ and $ \cmaxplus$, are used, e.g. for A* stack decoding to find best candidates~\cite{pau:92}. 

\item In ML, $\nnR$ itself is used to model uncertainty as probabilities and $\hR$ as log-probabilities. 
Although many of the problems and solutions cited above for AI can also be considered as part of ML, 
a recent branch of ML is solely  based upon the R\'enyi entropy with $\alpha=2$, $\rentropy[2]{P_X} = \srentropy[1]{P_X}$~\cite{pri:10}. 
%
Importantly, recall that every possible H\"older mean can be expressed as the arithmetic mean of a properly exponentiated kernel, whence  the importance of this particular R\'enyi entropy  would come. 

\item In CI, the sub-semifield obtained by the restriction of the operations to $\{\bot, e, \top\}$ appears as a \emph{ternary logic}---apart from the Boolean semifield, which is a sub-semifield of every complete semifield by restricting the carrier set to $\{\bot, \top\}$. This ternary subsemifields, as seen in Proposition~\ref{eq:smf:times:neg} and Theorem~\ref{theo:entropy:smf}.
Furthermore, Spohn's logical \emph{Rank theory}~\cite{spo:12} essentially leverages the isomorphims of semifields between $\nnR$ and the $\cminplus \equiv \hRr[\infty]$ in logical applications. 
Mathematical morphology and morphological processing need to operate in the dual pair $(\cmaxplus, \cminplus) \equiv (\hRr[-\infty], \hRr[\infty])$ for image processing applications~\cite{ron:90}. 
\end{itemize}

\end{enumerate}

\begin{Example}
Beyond rigid disciplinary boundaries, the Viterbi algorithm~\cite{vit:67} is the paramount example of a discovery that leads to non-linear, positive semifield algebra in an algorithmically-generic and application-independent setting~\cite{mar:17}. 
Initially devised as a teaching aid for convolutional codes~\cite{vit:67}, it was soon proven to be  an optimal algorithm for shortest-path decoding in a specific type of network~\cite{for:73}. But also, when this network comes from the unfolding over time of a Markov chain, it can be used to recover an ``optimal'' sequence of states and transition  over a generative model for a given sequence of observations~\cite{rab:89}. 
In this natural generalization, it has been applied to text and speech recognition and synthesis, among other cognitively-relevant applications that used to be considered part of ``classical'' AI but are modernly better tackled with more specialized ML or CI techniques. 

A crucial issue in this application was to realize that the ``optimality'' of the decoding strategy is brought about by the nature of the operations being used in the decoding---in the language of this paper, it is $\cminplus$-optimal. 
It follows that several other algorithms can be built with the template of the Viterbi algorithm by changing the underlying semifield, but all require that the addition be \emph{idempotent}, that is $\forall a \in S, a \oplus a = a$.
\qed
\end{Example}

These hints lead us to our \emph{main conjecture}, namely 
that \emph{applications in Machine Intelligence---whether AI, ML or CI---operate with information, equivalent probability or proxies thereof, 
and those calculations are, therefore, better conceptualized and successfully operated with the adequate dual pairs of positive semifields.
}



%
%
%
%


%

%
%
%
%
%


\section{Conclusions}
\label{sec:conc}
In the context of information measures, we have reviewed the notion of positive semifield---a positive semiring with a multiplicative group structure---distinct from that of the more usual fields with an additive group structure: in positive semirings there are no additive inverses, but there is a ``natural order'' compatible with addition and multiplication. 

Through Pap's $g$-calculus and Mesiar and Pap's semifield Construction, we have related the H\"older means 
to the shifted R\'enyi measures of information $\srentropy[r]{P_X}$ for pmf $P_X$, which appear as just the logarithm of the Kolmogorov-Nagumo means in different semifields obtained by ranging the $r$ parameter in $[-\infty, \infty]$. 
As a fundamental example, we provide the rewriting of the H\"older means in $\nnRr$ and its dual, which provides the basis for the shifted Renyi entropy, cross-entropy and divergence. 

Our avowed intention with this exploration was to 
provide a conjecture,  from an information theoretic point of view, about the abundance  of semifield valued quantities  in a variety of machine learning and computational intelligence tasks. 
Namely, 
that such semifield-valued quantities are being used either directly as R\'enyi information measures---including Shannon's---or indirectly as proxies of such.

\vspace{6pt} 




\authorcontributions{
Conceptualization, Francisco J Valverde-Albacete and Carmen Pel\'aez-Moreno; %
Formal analysis, Francisco J Valverde-Albacete and Carmen Pel\'aez-Moreno; 
Funding acquisition, Carmen Pel\'aez-Moreno; 
Investigation, Francisco J Valverde-Albacete and Carmen Pel\'aez-Moreno; 
Methodology, Francisco J Valverde-Albacete and Carmen Pel\'aez-Moreno; 
Writing – original draft, Francisco J Valverde-Albacete; 
Writing – review \& editing, Francisco J Valverde-Albacete and Carmen Pel\'aez-Moreno.
}

\funding{This research was funded by he Spanish Government-MinECo project
TEC2017-84395-P. 
}


\conflictofinterest{The authors declare no conflict of interest.}

\reftitle{References}



\externalbibliography{yes}
\bibliography{Smf4Entropy}

%
\iftoggle{graphicalSummary}{ %
\begin{figure}
\subfloat[Tranformation block $\overline Y = f(\overline X)$ with $(\overline X, \overline Y)\sim P_{\overline X \overline Y}$]{
	\resizebox{0.6\columnwidth}{!}{
	\input{transformationChain.tex}
	}
}
\subfloat[Entropy diagram for $P_{\overline X \overline Y}$%
]{
	\resizebox{0.4\columnwidth}{!}{
		\input{mod_idiagram_color_multivariate.tex}
		}
}
\\
 \subfloat[Schematic CMET for $P_{\overline X \overline Y}$ with formal interpretation.]{
 	\resizebox{0.6\linewidth}{!}{
  		\input{annotated_triangle_CMET_agg_formal.tex}
  		}
  	}
  		\subfloat[CMET of $\overline Y = f(\overline X)$  for ICA, PCA and log on Iris]{
		\includegraphics[width=0.5\linewidth]{featureSel_CMET_compare_PCA_ICA_iris.jpeg}
}

\caption{GRAPHICAL ABSTRACT
}
\end{figure}

}{}%
\end{document}